# Machine learning for prediction of extreme statistics in modulation instability

Mikko Närhi,[1,][*] Lauri Salmela,[1,][*] Juha Toivonen,[1]
Cyril Billet,[2] John M. Dudley,[2] and Goëry Genty[1]

[1]*Tampere University of Technology,*

*Laboratory of Photonics, FI-33101 Tampere, Finland*

[2]*Institut FEMTO-ST, Université Bourgogne Franche-Comté*

*CNRS UMR 6174, 25000 Besançon, France*

## Abstract

A central area of research in nonlinear science is the study of instabilities that drive the emergence of extreme events. Unfortunately, experimental techniques for measuring such phenomena often provide only partial characterization. For example, real-time studies of instabilities in nonlinear fibre optics frequently use only spectral data, precluding detailed predictions about the associated temporal properties. Here, we show how Machine Learning can overcome this limitation by predicting statistics for the maximum intensity of temporal peaks in modulation instability based only on spectral measurements. Specifically, we train a neural network based Machine Learning model to correlate spectral and temporal properties of optical fibre modulation instability using data from numerical simulations, and we then use this model to predict the temporal probability distribution based on high-dynamic range spectral data from experiments. These results open novel perspectives in all systems exhibiting chaos and instability where direct time-domain observations are difficult.

---

[*] These two authors contributed equally

## I. INTRODUCTION

A characteristic feature of many nonlinear dispersive systems is the process known as modulation instability (MI), which describes how noise on an input signal can be exponentially amplified to create localised structures of high intensity [1, 2]. There has been significant interest in studies of MI in nonlinear Schrödinger equation (NLSE) systems, with many experiments reported in fibre optics, hydrodynamics and other systems [3].

When seeded by noise, the localised structures emerging from MI show complex dynamics and random statistics, and it has even been suggested that MI may be linked to the development of extreme events or rogue waves [4–6]. Such studies have been of particular interest in the field of nonlinear fibre optics, because recent developments in real-time measurement techniques [7, 8] have allowed the emergent dynamics to be characterized experimentally in both the temporal and spectral domains. Specifically, in the temporal domain, although optical MI typically occurs on timescales that preclude direct electronic measurement, time-lens magnification has been used to characterize picosecond random breathers and solitons [9, 10]. In the spectral domain, the dispersive Fourier transform (DFT) has permitted real-time characterisation of a range of instabilities in both in optical fibres and laser cavities [11–17].

These new real-time measurement techniques have essentially revolutionized the study of ultrafast instabilities in nonlinear fibre optics [18–20], but they nonetheless remain limited in several important respects. For example, time lens magnification is experimentally complex, typically involving a nonlinear wavelength conversion process which constrains the measurement bandwidth and power. As a result, there are relatively few experiments that have directly measured ultrafast (picosecond) extreme events in the time domain [9, 10]. The DFT technique is experimentally simpler because it involves only propagation in dispersive fibre, but is typically associated with a low dynamic range of only 20-25 dB [21]. This a significant limitation to the detailed study of extreme events in MI which are associated with extension in the spectral wings below the -40 dB level [22, 23].

In this paper, we describe the development of a new high-dynamic range real time spectrometer that allows the analysis of unstable MI spectra with an experimental dynamic range approaching 60 dB. Although our measurements are performed only in the spectral domain, the application of Machine Learning to our spectral data allows us to nonetheless predict statistics for the maximum intensity of the localised temporal peaks in the MI field, peaks which are preferentially associated with rogue wave events. Our approach is based first on the use of numerical simulation data to train a Machine Learning model (based on a neural network) to correlate the spectral and temporal properties of MI. We then use the model to predict the temporal peak intensity probability distribution based on high-dynamic range experimental measurements of modulation instability in an optical fibre system. This analysis allows us to extract the statistics of the shot-to-shot intensity maxima, and obtain a predicted probability density function which is in excellent agreement with numerical modelling. Aside from the direct relevance of our results to optics, our approach has a far wider impact in showing how machine learning applied to only spectral data can still be successfully used to predict the properties of extreme events in the time domain.

## II. MODULATION INSTABILITY AND MACHINE LEARNING

Machine Learning is an umbrella term that describes the use of statistical techniques to analyse data sets with the aim of detecting patterns and building predictive models. Machine Learning has been widely applied to areas such as control systems, speech processing, neuroscience, and computer vision [24], and has very recently been applied to predicting the behavior of chaotic systems [25, 26]. Applications of Machine Learning in the field of photonics is also relatively recent, but a number of studies have been reported in laser optimization [27, 28], ultrashort pulses measurements [29], label-free cell classification [30], imaging [31–33], and coherent communications [34]. In our case, we aim to apply the techniques of Machine Learning to the study of chaotic nonlinear dynamics in optics, with the particular aim of predicting statistics for the maximum intensity of temporal peaks in modulation instability based only on spectral measurements.

Machine Learning first involves a training step, where a set of data with known characteristics is input into a model in order to determine a transfer function capable of correlating desired input and output properties. To this end, we used stochastic numerical simulations of the NLSE to generate a large ensemble of training data (both temporal and spectral) associated with a chaotic MI field. In particular, our simulations model our experiments (described below) where MI develops from picosecond pulses injected into the anomalous dispersion regime of an optical fibre. The simulations consider input pulses of 3 ps duration (full width at half maximum FWHM) and 175 W peak power evolving over a propagation distance of 0.68 m. To examine the effect of noise on shot-to-shot variations in the temporal and spectral properties, a broadband quantum-limited one photon-per-mode spectral noise background is included in the initial conditions [35]. Full details of the model used and all simulation parameters are found in the Methods section.

Typical results from a single simulation showing the evolution of spectral and temporal properties with distance are plotted in Fig. 1. We see MI characteristics showing the growth of distinct MI sidebands in the spectral domain (Fig. 1a) associated with the development of a strong temporal modulation and the development of localized peaks (breathers) on the pulse envelope (Fig. 1b).

In the picosecond regime, MI dynamics are highly sensitive to input noise, and for identical initial pulses but with a different random noise background, the spectral and temporal evolution can vary dramatically. This is shown explicitly in Fig. 1c and Fig. 1d where we plot 4 output spectral and temporal intensity profiles for different input noise, as well as the corresponding average profiles calculated over a larger number of 50,000 realisations. The 4 single-shot profiles clearly show complex structure and vary dramatically from shot-to-shot, but of course these unstable characteristics are not seen when the spectra and temporal profiles are averaged. It is for this reason that real-time measurement techniques have proven so valuable in understanding the nonlinear dynamics of MI.

Numerical simulations are extremely important in yielding insight into the statistics associated with the shot-to-shot variations of MI [36]. To this end, the solid line in Fig. 1e plots the probability density function (PDF) of the intensity of the localized MI peaks across the pulse envelope. This PDF is calculated from the $\sim 10^6$ temporal peaks identified from analysing the structure on the temporal envelopes obtained from the ensemble of 50,000 realizations. This probability distribution shows typical characteristics of MI with an extended tail, and the dashed vertical line shown in the tail region indicates the rogue wave threshold intensity $I_{\rm RW}$ defined as $I_{\rm RW} = 2I_{1/3}$ where $I_{1/3}$ is the mean intensity of the highest third of intensity peaks.

In the context of relating MI dynamics to the appearance of extreme events and rogue waves, our aim is to predict the intensity of the maximum peak occurring in a given temporal profile (i.e. the points indicated by circles in Fig. 1d) from only the corresponding spectral intensity profile. Note that the associated PDF of these maximum intensity peaks from the simulation data is shown as the red dashed line in Fig. 1e, and so it is clear that by focussing on the maximum peaks, we preferentially select out those events which have a greater probability to be classified as rogue wave events from the full distribution.

However, predicting the intensity of these temporal peaks from spectral profiles without the spectral phase is an extremely difficult problem because of the complexity of the MI temporal and spectral structure. Moreover from an experimental point of view, the highest temporal peaks are associated with broad exponentially decaying spectral wings extending of many 10's of dB dynamic range. And even determining the spectral bandwidth is not straightforward when dealing with noisy spectra consisting of multiple breathers with random amplitude and phase. As we will see, however, when combined with a novel experimental technique for high dynamic range spectral measurement, Machine Learning provides a robust and convenient solution that allows this problem to be solved.

The specific approach we use is based on a feed-forward neural network model to relate the input (spectral intensity profile) and output (temporal intensity maximum) obtained from stochastic MI simulations as illustrated in Fig 2. In particular, the spectral intensity from a single simulation realisation is written as a vector input $X = [x_1, x_2...x_N]$ where $x_i$ is the spectral intensity as wavelength $\lambda_i$ and mapped via a neural network to a scalar output $y$ corresponding to the maximum intensity of the associated temporal profile. The objective here is to use the training data to determine the weights and biases of the constituent nodes (neurons) that allow the network to perform as a transfer function to link $X$ and $y$. In our case, the neural network was trained using data from an ensemble of 30,000 simulations to generate spectral and temporal intensity profiles at the fibre output and, anticipating the use of this network on experimental data, the spectra were pre-processed to account for experimental conditions such as wavelength-dependent spectral response and system resolution (see Methods for full details).

After training, the model was tested on 20,000 simulations from a distinct ensemble of data not used in the training step. The aim here is to test how well the transfer function obtained from training is able to predict the maximum temporal intensity from a given simulated spectrum, by comparing the predicted value with the known value from the time-domain simulation data. The results of this test are shown in Fig. 3. Here Fig. 3a shows a false color density plot of the predicted maximum temporal intensity against the "target" value extracted from the simulation temporal data for the 20,000 simulation realisations.

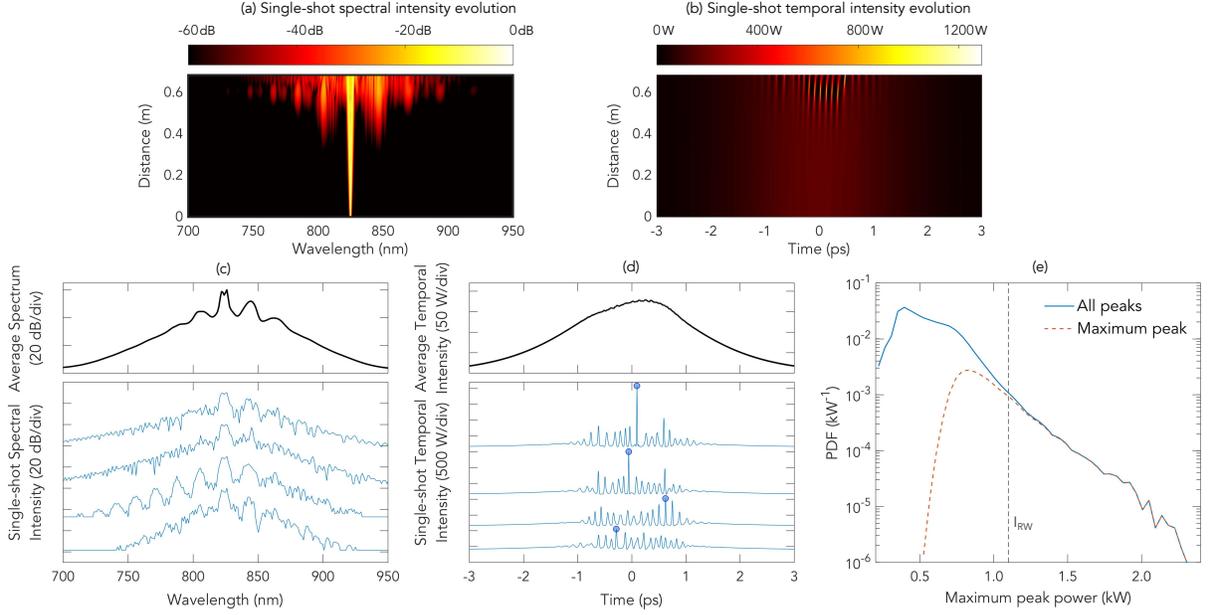

FIG. 1. Simulated MI dynamics from picosecond pulse propagation in optical fibre. (a) and (b) show results from single simulations illustrating spectral and temporal evolution over 0.68 m of propagation. (c) Spectral data from multiple simulations: the top figure shows the average spectrum over 50,000 realisations; the bottom figure shows the spectral output from 4 realisations to illustrate the complexity of the spectra and the shot-to-shot variations. (d) Temporal data from multiple simulations: the top figure shows the average temporal intensity over 50,000 realisations; the bottom figure shows the temporal intensity from the 4 realisations corresponding to (c) to illustrate the strong temporal modulation observed. The intensity peak in each case (shown by a circle) is the parameter we are aiming to predict from the corresponding spectral data. (e) Calculated probability density function of temporal intensity peaks from simulation data. The solid line shows results from all peaks (over a 1.5 ps window) while the dashed line shows only the distribution of the maxima temporal intensity peaks.

Here, in order to highlight the clustering of data points, the density plot uses a histogram representation where the data points are grouped into bins of constant area. The color scale shown corresponds to the normalised density of points in a particular bin. Note also the log scale for better visualization. We see clear clustering around the expected one-to-one linear relationship, with very strong correlation (Pearson correlation coefficient $\rho = 0.92$). It is also interesting to calculate the associated probability density distribution of the maximum temporal intensity, and this is shown in Fig. 3b. Here the probability density function from the simulation data (solid blue line) clearly shows the long tail previously observed in previous studies of NLSE instabilities [6]. The corresponding probability density function *predicted* by the Machine learning algorithm is shown as the red dashed line and it is clear that the algorithm performs impressively in predicting the shape of the probability distribution, especially the slope of the distribution tail as it extends to higher power events over nearly three orders of magnitude.

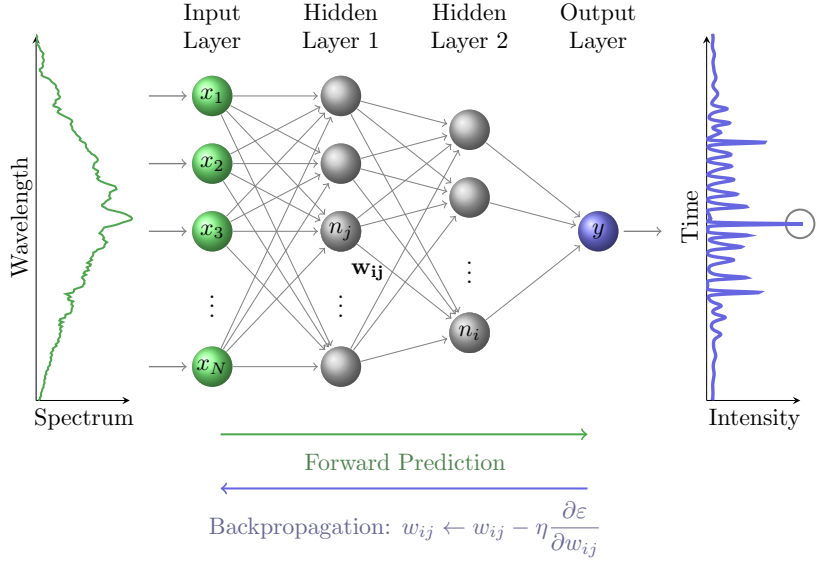

FIG. 2. Schematic of the neural network model used to correlate spectral and temporal characteristics of MI. A spectral intensity vector $X = [x_1, x_2...x_N]$ is input to a feed-forward neural network consisting of 2 hidden layers and a single output node $y$ corresponding to the maximum instantaneous peak power in the time-domain intensity profile (shown as the circled peak). The weights $w_{ij}$ of the network nodes correspond to the arrows connecting the node $n_i$ in layer $k$ to node $n_j$ in layer $k-1$ and they are adjusted during back-propagation towards the negative gradient of the error function $\varepsilon$ with step size $\eta$ (see Methods).

In contrast, as a further test during this evaluation phase, we examined whether lower-dynamic range spectral measurements (e.g. from conventional fibre-based DFT) could also be suitable for such Machine Learning analysis. To this end, we truncated the dynamic range of the 20,000 spectra from the test ensemble, applying a dynamic range of 25 dB which is typical for real time DFT systems, and we plot the corresponding 3D histogram results using this data in Fig. 3(c). From Fig. 3c it is clear that there is greatly reduced visual clustering around the one-to-one relationship and indeed the Pearson correlation coefficient here is only $\rho = 0.69$. Moreover, the predicted probability density function shown in Fig. 3(d) fails to reproduce the slope of the tail. We performed similar tests over a wider range of parameters, and found that Machine Learning was only able to construct a reliable predictive model for spectral data possessing a dynamic range exceeding 50 dB.

## III. EXPERIMENTAL SETUP AND RESULTS

The predictive model obtained from training was then applied to experimental measurements of noise-induced MI. The noisy MI field was generated by injecting pulses of duration 3 ps (FWHM) and peak power 175 W from a 80 MHz modelocked Ti:Sapphire laser at 825 nm into a 0.68 m length photonic crystal fibre with zero-dispersion wavelength around 750 nm. See Methods for full experimental detail, and note that the simulations described above used parameters identical to experiment. At the pump wavelength of 825 nm, the fibre used exhibits strong anomalous dispersion such that clear characteristics of MI are observed. To characterise the shot-to-shot fluctuations in the MI field, we developed a novel measurement setup capable of recording single-shot spectra with high dynamic range as shown in Fig. 4. This setup is described in full in the Methods section but the principle is based on first reducing the MI signal repetition rate to 150 kHz (using an acousto-optic modulator AOM) and then using a rapidly-rotating mirror to scan sequential output spectra onto different vertical positions of the entrance slit of a Czerny-Turner spectrograph with 1 nm resolution. In order to increase the dynamic range of the measurement, differential attenuation was used to capture the central region and the wings of the spectra separately and post-processing was then used to recombine the measured spectral components. With this technique, we obtained a dynamic range approaching 60 dB, a near four-order magnitude improvement compared to conventional fibre DFT.

Experimental results are shown in Fig. 5. In particular, Fig. 5a shows a sequence of 60 consecutive spectra to illustrate how the real-time technique is able to capture the large shot-to-shot fluctuations expected from MI in this picosecond regime [9, 12, 35]. It is especially significant that the high dynamic range reveals the variations in the structure of the spectral wings below the -40 dB level. As a check on the fidelity of these measurements, we computed an average spectrum from a larger sequence of 3,000 single-shot spectra, and these results are shown as the solid red line in Figure 5b. We compare these results with an independent measurement (solid yellow line) taken using an integrating optical spectrum analyzer (OSA) and it is clear there is very good agreement, although we note a discrepancy compared to the OSA for wavelengths beyond 875 nm due to reduced throughput efficiency of the system (i.e. grating and camera response). As a comparison with conventional fibre-DFT, the inset to Fig. 5 shows the average from our scanning real-time setup with that obtained using a standard fibre-DFT approach to highlight the near 4 orders of magnitude improvement obtained. As a further illustration of how our set-up allows us to measure significant shot-to-shot differences in the spectral wings, Fig. 5c compares the structure of two measured single-shot spectra (red solid line) with the computed average (gray dashed line).

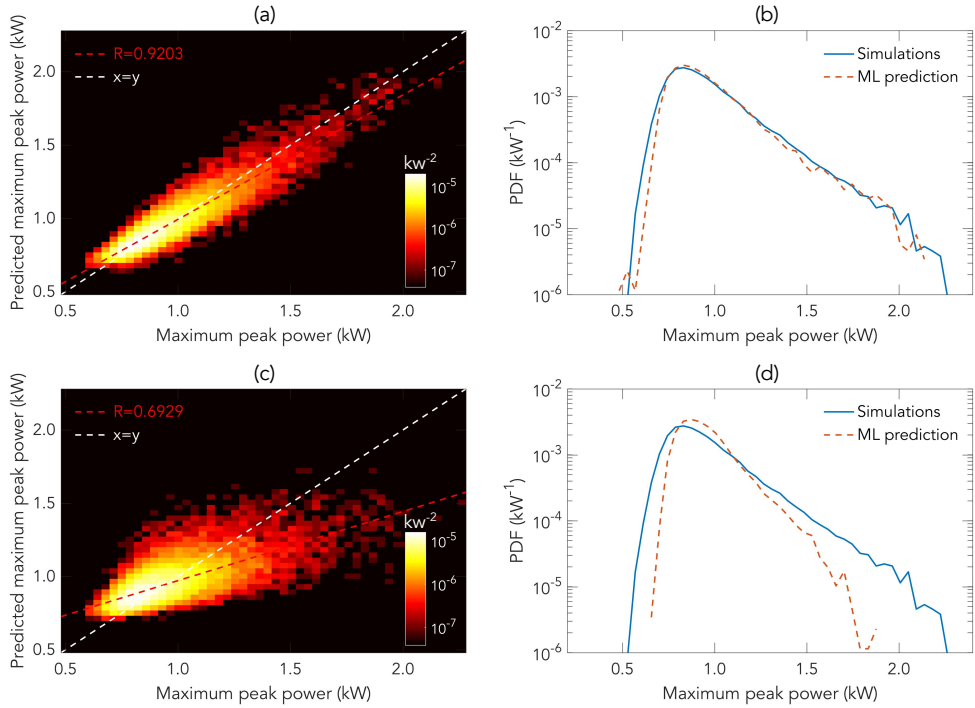

FIG. 3. (a,c) shows a comparison of the maximum intensity (instantaneous peak power) predicted by the machine learning algorithm with the exact value from the simulated time domain data. The results are shown as a false colour representation of a histogram which shows in logarithmic scale the normalized density of points grouped into bins of constant area. The dashed white line marks the 1-to-1 correspondence between the maximum intensity (instantaneous peak power) predicted by the machine learning algorithm with the exact value from the simulated time domain data. The value $R$ in the legend is the Pearson correlation coefficient. (b,d) Probability density function (PDF) of the maximum temporal intensities predicted by the machine learning algorithm (blue line) compared with the PDF calculated from the simulated time domain data (red line). An ensemble of 20,000 simulated single-shot spectra were used for the comparison. The top and bottom panels corresponds to truncating the input spectra data at 60 dB (except on the long wavelength side, see main text) and 25 dB dynamic range, respectively.

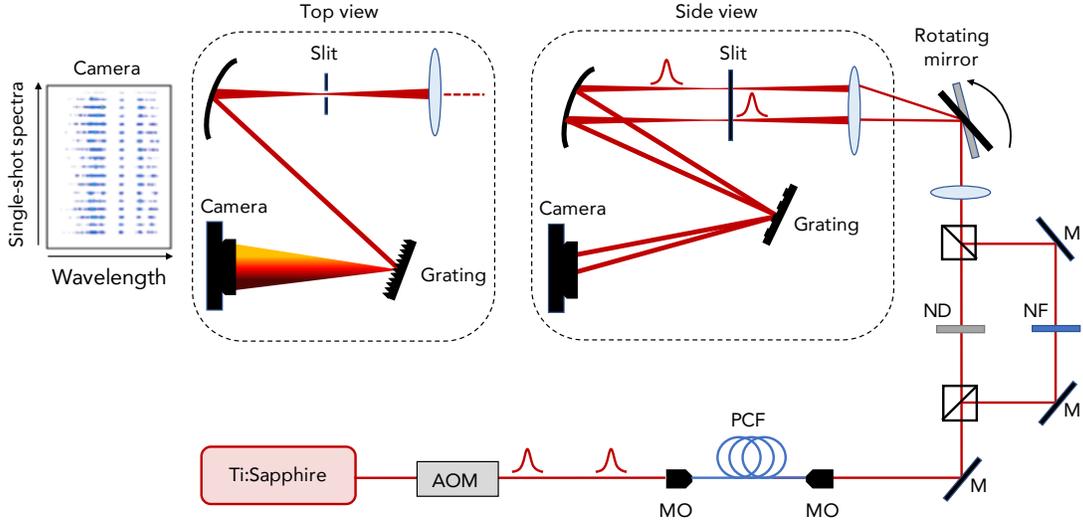

FIG. 4. Experimental setup. Ti:Sapphire: Titanium-Sapphire mode-locked laser, AOM: acousto-optic modulator, MO: microscope objective, PCF: photonic crystal fibre, M: mirror; ND: neutral density filter, NF: notch filter. Side views and top views of the grating imaging setup are shown.

As discussed above, we also performed numerical simulations for our experimental parameters, and for completeness in Fig. 5a, we show the average spectrum calculated from an ensemble of 30,000 such simulations (blue solid line). In this context we note that when single-shot simulation results were used in the Machine Learning training, simulated spectra were multiplied by a spectral response function to match the experimental fall off above 875 nm. This ensures that the transfer function obtained using simulation data can be applied to experimental results.

We next apply the Machine Learning transfer function to an ensemble of 3,000 experimental single-shot spectra, aiming to predict the maximum temporal intensity associated with each of the measured spectral intensity profiles. The results are shown in Fig. 6 where we compare the predicted probability density function from experimental data (dashed red) with that from numerical simulations (solid blue). We can see very good agreement between the experimental and simulation probability density functions, including in the slope of the distribution tail. These results show very clearly that even though the only available experimental data is that of the spectral intensity, we can nonetheless extract significant physical information about the corresponding temporal behaviour, and predict in particular the threshold value for extreme event classification.

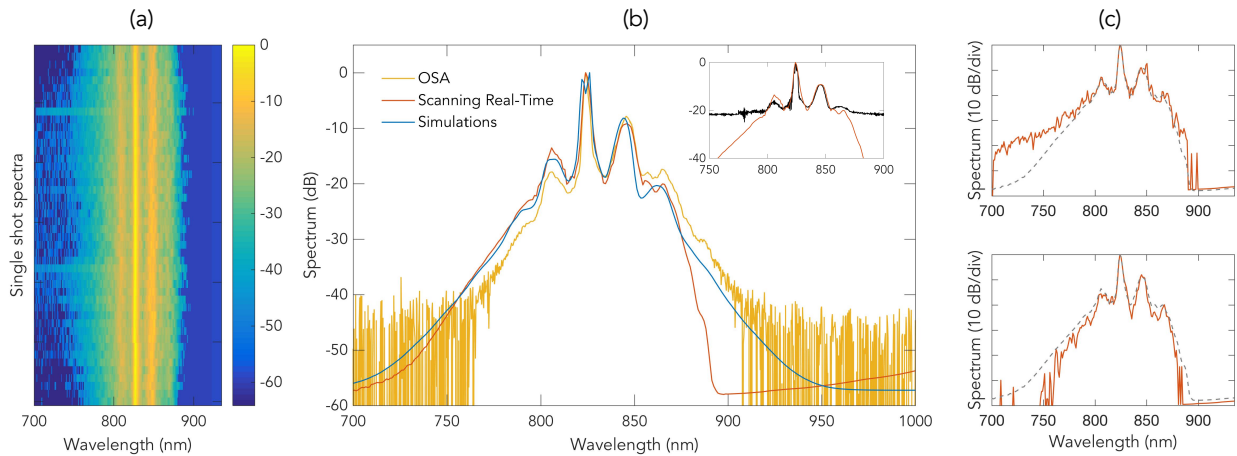

FIG. 5. (a) Series of 60 recorded spectra stacked along the vertical direction to illustrate the shot-to-shot fluctuations seen with MI. (b) Average spectrum measured with our scanning real-time technique (red), average spectrum measured with an optical spectrum analyzer (yellow), and simulated average spectrum (blue). The inset shows the comparison between the 60 dB dynamic range scanning spectral technique (red) and a conventional fibre-DFT with only $\sim 20$ dB dynamic range (black). (c) Two selected single-shot spectra from experiments (red) compared with the average experimental spectrum (dashed gray).

## IV. DISCUSSION

There are several major conclusions to be drawn from these results. Firstly, for the specific optical system studied here, we have shown that real-time measurements of only the spectral intensity can be combined with techniques from Machine Learning to yield quantitative information about temporal characteristics. In particular, the use of numerical simulations to train a predictive model allows temporal extreme events to be identified provided the real-time spectral measurement technique possesses sufficient dynamic range. In this regard, a further significant element of our results is the experimental technique used to capture shot-to-shot spectra with nearly 60 dB dynamic range. This approach is experimentally straightforward and can be implemented at all wavelengths where suitable spectrometers are available.

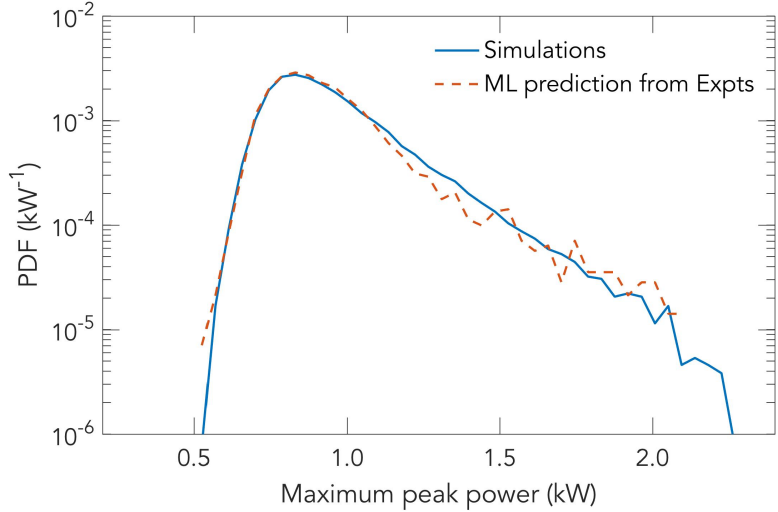

FIG. 6. Probability density function (PDF) of the maximum peak power of the temporal intensity profiles predicted by the machine learning algorithm from an ensemble of 3,000 single-shot experimental spectra (dashed red line) compared with the PDF calculated from the simulated time domain data (solid blue line).

Perhaps the most significant conclusion, however, arises from the fact that this technique is generic. Although demonstrated here using optical data, the principle of using Machine Learning to predict temporal behavior based only on spectral intensity measurements is extremely powerful. These results open novel perspectives in all physical systems exhibiting chaos and instability where direct time-domain observations are precluded.

## V. METHODS

### A. Numerical Modelling

Our numerical modelling is based on the well-known generalized NLSE model describing the evolution of a field envelope in an optical fibre [35]. Here, we model the propagation of 3 ps (FHWM) 175 W peak power hyperbolic-secant pulses in the anomalous dispersion regime of a 68 cm-long PCF (NKT Photonics NL-PM-750) with Taylor-series expansion dispersion coefficients at 825 nm $\beta_2 = -1.03 \times 10^{-26}$ s$^2$m$^{-1}$, $\beta_3 = 4.74 \times 10^{-41}$ s$^3$m$^{-1}$, $\beta_4 = 2.35 \times 10^{-56}$ s$^4$m$^{-1}$, $\beta_5 = -1.17 \times 10^{-70}$ s$^5$m$^{-1}$, and $\beta_6 = -9.07 \times 10^{-85}$ s$^6$m$^{-1}$. The nonlinear coefficient $\gamma = 0.1$ W$^{-1}$m$^{-1}$. For completeness, we also include the Raman and shock-terms in the model, but for our parameter regime, these had minor influence on the dynamics (although the Raman effect does lead to the observed MI sideband asymmetry.) Noise was included in the frequency domain in the form of a one photon per mode background (with random phase). For the machine learning training and testing we generated an ensemble of 50,000 numerical simulations corresponding to different input noise seeds. The simulations used 4096 grid points with a temporal window of 12 ps corresponding to an 83 GHz spectral resolution.

### B. Machine Learning and Neural Network

The input vector to our Machine Learning model consists of a vector $X = [x_1, x_2...x_N]$ of $N$ spectral intensity bins which are sequentially fed through a feed-forward neural network consisting of an input layer, two dense layers of hidden nodes with a nonlinear activation function, and an output layer with a single linear output node $y$. The two dense hidden layers have 30 and 10 nodes, respectively, consisting of weights for each connection from the previous layer plus an additional bias term. The output of a generic layer $\mathbf{h}_k \in \mathbb{R}^M$ is calculated from the weighted sum of the outputs on the previous layer $\mathbf{h}_{k-1} \in \mathbb{R}^D$ followed by a nonlinearity. Here, $k$ notes the index of the layer, and $M$ and $D$ are the dimension of the output vectors for layers $k$ and $k-1$, respectively. For the output layer, in our regressive model $M$ is one. The weighted sum for layer $k$ is calculated from:

$$\mathbf{g}_k = \mathbf{W}_k \mathbf{h}_{k-1} + \mathbf{b}_k, \quad (1)$$

where $\mathbf{W}_k \in \mathbb{R}^{M \times D}$ is a matrix of weights between the layers $k-1$ and $k$. The vector $\mathbf{b}_k$ contains the bias terms for each node in layer $k$. The weighted sum is then followed by a hyperbolic tangent sigmoid activation function $f$:

$$\mathbf{h}_k = f(\mathbf{g}_k), \qquad (2)$$

producing the output of layer $k$. For the output layer, a linear activation function was used. Conjugate gradient back-propagation with Fletcher-Reeves updates [37, 38] was selected as the training function which determines how the weights and biases are adjusted. The mean square error function was used as the cost function. The training of the network involves two steps: forward and backward pass. First, the forward pass feeds one or more samples (i.e. spectral intensity vectors $X$) to the network. The backward pass then adjusts the weights and biases minimizing the cost function $\epsilon$. Similarly to other conjugate gradient methods, the weights $w_{ij}$ are iteratively adjusted by an amount $\Delta w_{ij}$ with learning rate $\eta$ towards the negative gradient of the cost function such that $\Delta w_{ij} = -\eta \frac{\partial \varepsilon}{\partial w_{ij}}$. This is then repeated with more samples being fed to the network and the weights and biases adjusted again during back propagation. This process leads to a network suitably configured after training to perform as the desired transfer function linking $X$ and $y$.

Because our goal is to apply the machine learning algorithm trained from the numerical simulations to real-world experimental data, the input spectral intensity vector was pre-processed so as to match that of the experimentally recorded single-shot spectral intensity measurements in aspects such as resolution, bandwidth, and wavelength grid. Specifically, the simulated spectra were convolved with a 1 nm (full-width half maximum) Gaussian function and interpolated onto the experimental wavelength grid. The spectral content for the training data was further limited from 705 nm to 875 nm that corresponds to the experimental noise floor on the short wavelengths side and decreased spectral efficiency of the instrument response on the long wavelength sides. With this pre-processing, the input vector then consists of 121 spectral intensity bins. The neural network was trained for 300 epochs using 30,000 simulated single-shot spectra of a noisy MI field. Furthermore, in order to provide the machine learning algorithm a more versatile training set that allows finding a more general pattern for estimating the maximum intensity in the time domain and thereby mitigating the effects of our experimental uncertainties, we used an additional ±5% random variation of the pulse peak power in the simulations.

### C. Scanning real-time spectral measurement setup

Single-shot MI spectra were measured in real-time at the fibre output using a rapidly rotating mirror mounted on a galvanometer (Nutfield QS-12) with angular speed $\omega = 240$ rev./min, and focused with a lens of focal length $f = 150$ mm at the entrance slit of a Czerny-Turner spectrograph. The spectrograph used a grating with 300 lines/mm and 500 nm blaze (ThorLabs GR25-0305) to disperse consecutive spectra onto different lines of a high-sensitivity electron-multiplying charged-coupled device (EMCCD) camera (Andor iXon 3), allowing single-shot spectral intensity measurements with a 1 nm resolution. With this scan rate and our setup, it was necessary to reduce the repetition rate of the laser to 150 kHz using an acousto-optic modulator, but acquisition speeds up to the MHz range would be possible either using a faster galvanometer, using a multi-pass geometry [39] or by increasing the focal length at the spectrograph entrance slit.

The camera was cooled to -80ºC and used 5× pre-amplifier gain to decrease the noise level to a single electron level corresponding to a maximum dynamic range close to 40 dB. In order to increase the effective dynamic range of the measurement, we used a differential spectral attenuation scheme that captures the central part and wings of the MI spectra separately with the same dynamic range. In this scheme, the MI field at the fibre output is divided between two arms of unequal length corresponding to a 200 ps delay. Differential attenuation was induced in the two arms using a notch filter with a 40 dB, 20 nm rejection band centred at 825 nm (Edmund Optics) and a variable neutral density filter, respectively. Beams from the two arms are then recombined with a beamsplitter such that the central part and wings of the individual spectra are recorded with the same dynamic range and 200 ps delay by the individual lines of the EMCCD. The spectral response of the system was carefully calibrated by measuring the mean spectrum with and without the filters. The full spectra are subsequently recombined by post-processing with an effective 60 dB dynamic range, representing a more than three-order of magnitude improvement compared to a conventional fibre-based DFT approach [12, 13]. Direct comparison of the average MI spectrum at the PCF output was performed with an integrating optical spectrum analyzer (Ando AQ6315B).

### D. DFT measurement setup

The conventional fibre DFT implemented for comparison with the scanning real-time approach used a 100 m custom fabricated fibre (IXfibre IXF-SM series) designed to be single-mode over a broad wavelength range in the near-infrared and with total dispersion $\beta_2 L = +4030\,\text{ps}^2$ at 825 nm. The input to the dispersive stretching fibre was attenuated to ensure linear propagation. The real-time spectra were recorded with a 25 GHz InGaAs photodiode (UPD-15-IR2-FC Alphalas) and 20 GHz real-time oscilloscope (DSA72004 Tektronix), leading to an effective resolution of $\sim 0.03$nm.

**FUNDING INFORMATION**

MN acknowledges the support from Kaute foundation and TUT graduate school. JMD acknowldges support from the French Investissements d'Avenir program, project ISITE-BFC (contract ANR-15-IDEX-0003). GG acknowledges the support from the Academy of Finland (grants 298463 and 318082).

**ACKNOWLEDGEMENTS**

D. Brunner, P. Ryczkowski and T. Sylvestre are acknowledged for useful discussions.